\documentclass[aps,prb,twocolumn,showpacs,amsmath]{revtex4}
\usepackage{latexsym}
\usepackage{amssymb}
\usepackage{graphicx}
\usepackage{bm}% bold math
\usepackage[english]{babel}
\usepackage[latin1]{inputenc}
\usepackage{amsmath}
\usepackage{amsfonts}
\newcommand{\pdag}{{\phantom{\dagger}}}
\newcommand{\bq}{\begin{equation}}
\newcommand{\eq}{\end{equation}}
\newcommand{\bn}{\begin{eqnarray}}
\newcommand{\en}{\end{eqnarray}}

\begin{document}

\title{Vibration-mediated resonant tunneling and shot noise through a %%@
molecular quantum dot}

\author{X. Y. Shen}
\affiliation{Department of Physics, Shanghai Jiaotong University,
1954 Huashan Road, Shanghai 200030, China}

\author{Bing Dong}
%\altaffiliation{Author to whom correspondence should be addressed. %%@
%Email:bdong@sjtu.edu.cn}
\thanks{Author to whom correspondence should be addressed. %%@
Email:bdong@sjtu.edu.cn.}
\affiliation{Department of Physics, Shanghai Jiaotong University,
1954 Huashan Road, Shanghai 200030, China}

\author{X. L. Lei}
\affiliation{Department of Physics, Shanghai Jiaotong University,
1954 Huashan Road, Shanghai 200030, China}

\author{N. J. M. Horing}
\affiliation{Department of Physics and Engineering Physics, Stevens Institute %%@
of Technology, Hoboken, New Jersey 07030, USA}

\begin{abstract}

Motivated by a recent experiment on nonlinear tunneling in a suspended Carbon %%@
nanotube connected to two normal electrodes [S. Sapmaz, {\it et al}., Phys. %%@
Rev. Lett. {\bf 96}, 26801 (2006)], we investigate nonequilibrium %%@
vibration-mediated sequential tunneling through a molecular quantum dot with %%@
two electronic orbitals asymmetrically coupled to two electrodes and strongly %%@
interacting with an internal vibrational mode, which is itself weakly coupled %%@
to a dissipative phonon bath. For this purpose, we establish rate equations %%@
using a generic quantum Langevin equation approach. Based on these equations, %%@
we study in detail the current-voltage characteristics and zero-frequency shot %%@
noise, paying special attention to the advanced or postponed of the appearance %%@
of negative differential conductance and super-Poissonian current noise %%@
resulting from electron-phonon-coupling induced {\em selective unidirectional %%@
cascades of single-electron transitions}.     

\end{abstract}

\date{\today}


\maketitle

\section{Introduction}

Recent progress in nanotechnology has facilitated the fabrication of %%@
single-electron tunneling devices using organic molecules. In turn, this has %%@
given rise to a large body of 
experimental\cite{jPark,hPark,Zhitenev,Yu,Pasupathy,LeRoy,Sapmaz} and %%@
theoretical\cite{Bose,Alexandrov,McCarthy,Mitra,Koch,Koch1,Koch2,Zazunov,Nowac%%@
%%@
k,Wegewijs,Haupt} work concerning phonon-mediated resonant tunneling through a %%@
quantum dot (QD) with strong coupling to an internal vibrational (phonon) mode %%@
(IVM).
In particular, Carbon nanotubes (CNT) have recently become the focus of much %%@
research interest because electronic transport measurements show that they %%@
demonstrate perfect signatures of phonon-mediated tunneling, e.g. stepwise %%@
structures in the current-voltage characteristics having equal widths in %%@
voltage and gradual height reduction by the Franck-Condon (FC) %%@
factor.\cite{LeRoy,Sapmaz}  

More interestingly, the experimental measurement of S. Sapmaz, {\it et %%@
al}.\cite{Sapmaz} has revealed some more subtle transport features in a %%@
suspended CNT connected to two electrodes, particularly the appearance of a %%@
weak negative differential conductance (NDC) at the onset of each phonon step %%@
followed by a sudden suppression of current (a strong NDC) at a finite bias %%@
voltage after several steps for a long CNT sample. Theoretically, the reason %%@
for the weak NDC is quite clear in that it is ascribed to the combined effect %%@
of strong electron-phonon coupling (EPC) and low relaxation, i.e. an %%@
unequilibrated phonon (hot phonon), and strongly asymmetric tunnel-couplings %%@
to the left and right electrodes.\cite{Bose,Koch2,Zazunov} Furthermore, %%@
McCarthy, {\it et al}. have theoretically predicted the catastrophic current %%@
decrease, but their calculations are based on the presumption that the EPC is %%@
dependent on the applied bias voltage. It is appropriate to explore yet %%@
another explanation of the origin of the strong NDC without such an %%@
assumption.

Up to now, most theoretical works have focused on the case of a single level %%@
coupling to the phonon mode. However, it is believed that this catastrophic %%@
current decrease is intimately related to the tunneling processes involving %%@
{\em two} electronic energy levels. Nowack and Wegewijs have considered %%@
vibration-mediated tunneling through a two-level QD with asymmetric couplings %%@
to an IVM.\cite{Nowack} Albeit their results display strong NDC through a %%@
competition between different FC tailored tunneling processes of the two %%@
levels, their calculations do not fully resolve the overall features of the %%@
experimental data in Ref.~\onlinecite{Sapmaz}. Following a suggestion by %%@
Hettler, {\it et al}.,\cite{Hettler} that the interplay of strong Coulomb %%@
blockade and asymmetric tunnel-couplings of two levels to electrodes can lead %%@
to a strong NDC under certain conditions without EPC, we employ a model with %%@
two electronic orbitals having asymmetric couplings to the electrodes and with %%@
both strongly interacting with an IVM. We establish approximate rate equations %%@
at high temperature to describe resonant tunneling incorporating the %%@
unequilibrated phonon effect. Our results are in good qualitative agreement %%@
with the experiment data of Ref.~\onlinecite{Sapmaz}: in the weak bias voltage %%@
region where the first molecular orbital (MO) is dominant in electronic %%@
tunneling, the current shows the FC-type steplike structure and weak NDC at %%@
the onset of each step; while with increasing bias voltage the second MO %%@
becomes dominant in tunneling, inducing a sudden rapid reduction of current %%@
whether there is EPC or not. Furthermore, our studies predict that the onset %%@
of strong NDC as a function of voltage may differ from the value at which the %%@
second MO becomes active in resonant tunneling as described in %%@
Ref.~\onlinecite{Hettler} due to the unequilibrated phonon effect. We ascribe %%@
this to the EPC induced selective cascades of single-electron transitions %%@
between different charge states.      
In addition, we also analyze the current noise properties of this system. We %%@
find that the occurrence of weak and strong NDCs is accompanied by %%@
corresponding weak and strong enhancements of zero-frequency shot noise, %%@
respectively.

To emphasize the external-voltage-driven unequilibrated phonon effect in %%@
electronic tunneling in a CNT, we consider the IVM in our model coupled to an %%@
external dissipative environment (a phonon bath),\cite{Haupt} and we %%@
incorporate the dissipation mechanism of the unequilibrated phonon into the %%@
ensuing rate equations on a microscopic basis. For extremely strong %%@
dissipation, our results naturally reduce to those of equilibrated %%@
phonon-mediated tunneling. Therefore, our rate equations provide a valuable %%@
theoretical framework for analysis of the mechanism by which finite phonon %%@
relaxation influences the electronic transport properties of CNT. 

The outline of the paper is as follows. In Sec.~II, we describe the model %%@
system that we study, and rewrite the model Hamiltonian in terms of an %%@
electron-phonon direct product (EPDP) state representation, which is suitable %%@
for the ensuing theoretical derivation. 
In Sec.~III, we derive a set of rate equations using a generic quantum %%@
Langevin equation approach with a Markovian approximation, which facilitates %%@
investigation of uneqilibrated phonon and phonon dissipation effects on %%@
phonon-mediated electronic tunneling. In this section, we also derive the %%@
current formula using linear-response theory. In Sec.~IV, we describe %%@
MacDonald's formula for calculating zero-frequency shot noise by rewriting the %%@
ensuing rate equations in a number-resolved form.  
Then, we investigate in detail the vibration-mediated transport and shot noise %%@
properties of a molecular QD with two MOs in Sec.~V. Finally, a brief summary %%@
is given in Sec.~VI.

\section{Model Hamiltonian}

In this paper, we consider a generic model for a molecular QD (in particular %%@
the suspended CNT) with two spinless levels, one as the highest-occupied MO %%@
(HOMO) $\varepsilon_1$ and the other as the lowest-unoccupied MO (LUMO) %%@
$\varepsilon_2$, coupled to two electrodes left (L) and right (R), and also %%@
linearly coupled to an IVM of the molecule having frequency $\omega_0$ with %%@
respective coupling strengths $\lambda_1$ and $\lambda_2$. We suppose that %%@
this single phonon mode is coupled to a dissipative environment represented as %%@
a set of independent harmonic oscillators (phonon-bath). The model Hamiltonian %%@
is
\begin{subequations}
\bq
H = H_{leads}+H_{mol}+H_B+H_{I}, \label{hamiltonian}
\eq
with
\bn
H_{leads} &=& \sum_{\eta, {\bf k}} \varepsilon_{\eta {\bf k}} c_{\eta {\bf %%@
k}}^\dagger c_{\eta {\bf k}}^\pdag, \\
H_{mol} &=& \sum_{j=1,2} \varepsilon_j c_j^{\dagger} c_j^\pdag +U n_1 n_2 \cr
&& + \omega_0 a^\dagger a + \sum_{j=1,2} \lambda_j c_j^\dagger c_j^\pdag %%@
(a^\dagger + a), \label{Hmol} \\
H_B &=& \sum_{p} \omega_p b_p^\dagger b_p^\pdag, \\
H_{I} &=& H_T+H_{vB}, \\
H_T &=& \sum_{\eta,{\bf k},j} (V_{\eta j} c_{\eta {\bf k}}^\dagger c_j + {\rm %%@
H.c.}), \\
H_{vB} &=& (a^\dagger +a) \sum_{p} \kappa_p (b_{p}^\dagger + b_{p}),  
\en
\end{subequations}
where $c_{\eta{\bf k}}^\dagger$ ($c_{\eta{\bf k}}$) is the creation %%@
(annihilation) operator of an electron with momentum ${\bf k}$, and energy %%@
$\varepsilon_{\eta {\bf k}}$ in lead $\eta$ ($\eta=L,R$), and $c_{j}^\dagger$ %%@
($c_j$) is the corresponding operator for a spinless electron in the $j$th %%@
level of the QD ($j=1,2$). $U$ denotes interdot Coulomb interaction and %%@
$n_j=c_j^\dagger c_j^\pdag$ is the electron number operator in level $j$. %%@
$a^\dagger$ ($a$) and $b_p^\dagger$ ($b_p$) are phonon creation (annihilation) %%@
operators for the IVM and phonon-bath (energy quanta $\omega_0$, $\omega_p$), %%@
respectively. $\lambda_j$ represents the coupling constant between electron in %%@
dot $j$ and the IVM; $\kappa_p$ is the coupling strength between the IVM and %%@
phonon bath; $V_{\eta j}$ describes the tunnel-coupling between electron level %%@
$j$ and lead $\eta$.
Here, we denote the density of states of the phonon bath with respect to %%@
frequency $\omega_p$ by $D(\omega_p)$, and define the corresponding spectral %%@
density $J_B(\omega_p)$ as
\bq
J_B(\omega_p)=\kappa_p^2 D(\omega_p).
\eq
In the literature, the following form of the spectral density is usually %%@
considered:\cite{Grabert}
\bq
J_B(\omega_p)=\kappa_0 \omega_p^\alpha \theta(\omega_p), \label{pbspectral}
\eq
in which $\theta(x)$ is the Heaviside step function. The bath is said to be %%@
Ohmic, sub-Ohmic, and super-Ohmic if $\alpha=1$, $\alpha<1$, and $\alpha>1$, %%@
respectively. We use units with $\hbar=k_B=e=1$ throughout the paper. In %%@
addition, it is worth noting that we do not consider the EPC-induced coupling %%@
between two MO's in the Hamiltonian, Eq.~(\ref{Hmol}), because it is much %%@
weaker than the coupling of a single MO.

It is well-known that, in the strong electron-phonon interaction problem, it %%@
is very convenient to introduce a standard canonical %%@
transformation\cite{Mahan} to the Hamiltonian Eq.~(\ref{hamiltonian}), 
$\widetilde{H}=e^{S} H e^{-S}$, with $S=(g_1 n_1 + g_2 n_2) (a^\dagger - a)$ %%@
($g_j=\lambda_j/\omega_0$), which leads to a transformed Hamiltonian
\begin{subequations}
\bn
\widetilde{H}_{mol} &=& \sum_{j} \widetilde{\varepsilon}_j c_j^\dagger c_j + %%@
\widetilde{U} n_1 n_2 + \omega_0 a^\dagger a, \\
\widetilde H_{I} &=& \widetilde H_T+ \widetilde H_{vB}, \\
\widetilde{H}_{T} &=& \sum_{\eta,{\bf k},j} (V_{\eta j} c_{\eta {\bf %%@
k}}^\dagger c_j X_j + {\rm H.c.}), \\
\widetilde{H}_{vB} &=& (a^\dagger +a - 2 \sum_{j} g_j n_j) \sum_{p} \kappa_p %%@
(b_{p}^\dagger + b_{p}) ,  
\en
with $\widetilde{\varepsilon}_j = \varepsilon_j - %%@
\frac{\lambda_j^2}{\omega_0}$ and 
$\widetilde{U} = U- 2\frac{\lambda_1 \lambda_2}{\omega_0}$.
Importantly, the $X$-operator describes the phonon renormalization of dot-lead %%@
tunneling, 
\bq
X_j = e^{g_j (a-a^\dagger)}.
\eq
\end{subequations}

Obviously, the model described by the above Hamiltonian %%@
Eq.~(\ref{hamiltonian}) involves a many-body problem with phonon generation %%@
and annihilation when an electron tunnels through the central region. %%@
Therefore, one can expand the electron states in the dot in terms of direct %%@
product states composed of single-electron states and $n$-phonon Fock states.
In this two-level QD system, there are a total of four possible electronic %%@
states: (1) the two levels are both empty, $|0\rangle_{1} |0\rangle_{2} $, and %%@
its energy is zero; (2) the HOMO is singly occupied by an electron, $|1 %%@
\rangle_{1} |0\rangle_{2}$, and its energy is $\varepsilon_{1}$; (3) the LUMO %%@
is singly occupied, $|0\rangle_{1} |1\rangle_{2}$, and its energy is %%@
$\varepsilon_2$; and (4) both orbitals are occupied, $|1\rangle_{1} %%@
|1\rangle_{2}$, and its energy is $\varepsilon_1+\varepsilon_2+U$. Of course, %%@
if the interdot Coulomb repulsion is assumed to be infinite, the %%@
double-occupation is prohibited. For the sake of convenience, we assign these %%@
Dirac bracket structures as dyadic kets (bras):\cite{dyadic} namely, the %%@
slave-boson kets $e_n^{\dagger}=|0\rangle_{1} |0\rangle_{2}\otimes |n %%@
\rangle$, $d_n^{\dagger}=|1\rangle_{1} |1\rangle_{2}\otimes |n \rangle$, and %%@
pseudo-fermion kets $f_{1n}^{\dagger}=|1\rangle_{1} |0\rangle_{2}\otimes |n %%@
\rangle$, $f_{2n}^{\dagger}=|0\rangle_{1} |1\rangle_{2}\otimes |n \rangle$. %%@
Correspondingly, the electron operator $c_{j}$ and phonon operator $a$ can be %%@
written in terms of such ket-bra dyadics as:
\bn
c_{j} &=& \sum_{n=0}^{\infty} (e_n^\dagger f_{jn} + s_{\bar{j}} f_{\bar %%@
{j}n}^\dagger d_n), \\
a &=& \sum_{n=0}^{\infty} \sqrt{n+1} (e_{n}^\dagger e_{n+1}^\pdag + \sum_j %%@
f_{j n}^\dagger f_{j n+1}^\pdag + d_n^\dagger d_{n+1}^\pdag),
\en
with $\bar j\neq j$ and $s_{1(2)}=\pm 1$ ($s_2=-1$ is due to anti-commutation %%@
relation of the Fermion operator). With these direct product states and %%@
dyadics considered as the basis, the density-matrix elements may be expressed %%@
as $\rho_{00}^n=\langle \hat\rho_{00}^n \rangle=\langle e_n^{\dagger} %%@
e_n^\pdag \rangle$, $\rho_{jj}^n=\langle \hat\rho_{jj}^n\rangle=\langle %%@
f_{jn}^{\dagger} f_{jn}^{\pdag} \rangle$, and $\rho_{dd}^n=\langle %%@
\hat\rho_{dd}^n \rangle=\langle d_n^{\dagger} d_n^\pdag \rangle$. 

The transformed Hamiltonian can be replaced by the following form in the %%@
auxiliary particle representation:
\begin{subequations}
\bn
\widetilde{H}_{mol} &=& \sum_{n} [n\omega_0 e_n^\dagger e_n^\pdag + \sum_{j} %%@
(\widetilde{\varepsilon}_j +n\omega_0) f_{jn}^\dagger f_{jn}^\pdag \cr
&& + (\widetilde{\varepsilon}_1+\widetilde{\varepsilon}_2+\widetilde{U}+ %%@
n\omega_0) d_n^\dagger d_n^\pdag ], \\
\widetilde{H}_{T} &=& \sum_{\eta,{\bf k},j,n} [V_{\eta j} c_{\eta {\bf %%@
k}}^\dagger (e_n^\dagger f_{jn} + s_{\bar{j}} f_{\bar {j}n}^\dagger d_n) X_j + %%@
{\rm H.c.}], \cr
&& \\
\widetilde{H}_{vB} &=& \sum_{p} \kappa_p (b_{p}^\dagger + b_{-p}) \left [ %%@
\sum_{n} \sqrt{n+1} (e_{n}^\dagger e_{n+1}^\pdag \right. \cr
&& + \sum_j f_{jn}^\dagger f_{jn+1}^\pdag + d_{n}^\dagger d_{n+1}^\pdag +{\rm %%@
H.c.}) \cr
&& \left. - 2 \sum_{j,n} g_j (f_{jn}^\dagger f_{jn}^\pdag + d_{n}^\dagger %%@
d_{n}^\pdag) \right ].  
\en
\end{subequations}
Based on this transformed Hamiltonian, we derive a rate equation for %%@
description of the dynamics of the reduced density matrix of the combined %%@
electron and IVM system in the sequential tunneling regime.

\section{Quantum Langevin equation approach and rate equations}

In this analysis, we employ a generic quantum Langevin equation %%@
approach,\cite{Schwinger,Ackerhalt,Cohen,Milonni,Gardiner,Dong2,Dong3} %%@
starting from the Heisenberg equations of motion (EOMs) for the density-matrix %%@
operators $\hat\rho_{00}^n$, $\hat\rho_{jj}^n$ ($j=\{1,2 \}$), and %%@
$\hat\rho_{dd}^n$:
\begin{subequations}
\bn
i\dot{\hat\rho}_{00}^n &=& [e_n^\dagger e_n^\pdag, \widetilde{H}]_- = %%@
\sum_{\eta,{\bf k},j,m} [V_{\eta j} c_{\eta {\bf k}}^\dagger e_n^\dagger %%@
f_{jm} X_{j,nm} \cr
&& - f_{jm}^\dagger e_n X_{j,nm}^\dagger c_{\eta {\bf k}}] + \sum_{p} \kappa_p %%@
(b_{p}^\dagger + b_{-p}) \cr
&& \times (\sqrt{n+1} e_n^\dagger e_{n+1}^\pdag + \sqrt{n} e_n^\dagger %%@
e_{n-1}^\pdag - {\rm H.c.}), \label{eom:r00}\\
i\dot{\hat\rho}_{11}^n &=& [f_{1n}^\dagger f_{1n}^\pdag, \widetilde{H}]_- = %%@
\sum_{\eta,{\bf k},m} [c_{\eta {\bf k}}^\dagger (V_{\eta 2} f_{1n}^\dagger  %%@
d_{m} X_{2,nm} \cr
&& - V_{\eta 1} e_{m}^\dagger f_{1n} X_{1,mn}) + (V_{\eta 1} f_{1n}^\dagger %%@
e_m X_{1,mn}^\dagger \cr
&& - V_{\eta 2} d_m^\dagger f_{1n} X_{2,nm}^\dagger) c_{\eta {\bf k}}] + %%@
\sum_{p} \kappa_p (b_{p}^\dagger + b_{-p}) \cr
&& \times (\sqrt{n+1} f_{1n}^\dagger f_{1n+1}^\pdag + \sqrt{n} f_{1n}^\dagger %%@
f_{1n-1}^\pdag - {\rm H.c.}), \cr
&& \\
i\dot{\hat\rho}_{22}^n &=& [f_{2n}^\dagger f_{2n}^\pdag, \widetilde{H}]_- = %%@
\sum_{\eta,{\bf k},m} [- c_{\eta {\bf k}}^\dagger (V_{\eta 1} f_{2n}^\dagger  %%@
d_{m} X_{1,nm} \cr
&& + V_{\eta 2} e_{m}^\dagger f_{2n} X_{2,mn}) + (V_{\eta 1} f_{2n}^\dagger %%@
e_m X_{2,mn}^\dagger \cr
&& + V_{\eta 2} d_m^\dagger f_{2n} X_{1,nm}^\dagger) c_{\eta {\bf k}}] + %%@
\sum_{p} \kappa_p (b_{p}^\dagger + b_{-p}) \cr
&& \times (\sqrt{n+1} f_{2n}^\dagger f_{2n+1}^\pdag + \sqrt{n} f_{2n}^\dagger %%@
f_{2n-1}^\pdag - {\rm H.c.}), \cr
&& \\
i\dot{\hat\rho}_{dd}^n &=& [d_n^\dagger d_n^\pdag, \widetilde{H}]_- = 
\sum_{\eta,{\bf k},m} [c_{\eta {\bf k}}^\dagger (V_{\eta 1} f_{2m}^\dagger  %%@
d_{n} X_{1,mn} \cr
&& - V_{\eta 2} f_{1m}^\dagger d_{n} X_{2,mn}) - (V_{\eta 1} d_{n}^\dagger %%@
f_{2m} X_{1,mn}^\dagger \cr
&& - V_{\eta 2} d_n^\dagger f_{1m} X_{2,mn}^\dagger) c_{\eta {\bf k}}] + %%@
\sum_{p} \kappa_p (b_{p}^\dagger + b_{-p}) \cr
&& \times (\sqrt{n+1} d_n^\dagger d_{n+1}^\pdag + \sqrt{n} d_n^\dagger %%@
d_{n-1}^\pdag - {\rm H.c.}), \label{eom:rdd}
\en
where
\bn
X_{j,nm} &=& \langle n |X_{j}(t) |m\rangle,\\
X_{j,nm}^\dagger &=& \langle m |X_{j}^\dagger(t) |n\rangle.
\en
\end{subequations}
These matrix elements can be calculated as:\cite{Mahan}
\bn
X_{j,nm} &=& X_{j,nm}^\dagger \cr
&=& \left \{
\begin{array}{cc}
e^{-g_j^2/2} g_j^{m-n} \sqrt{\frac{n!}{m!}} L_n^{m-n} (g_j^2), & n\leq m, \\
e^{-g_j^2/2} (-g_j)^{n-m} \sqrt{\frac{m!}{n!}} L_m^{n-m} (g_j^2), & n> m,
\end{array}
\right.
\en
where $L_n^m(x)$ is the generalized Laguerre polynomial.
The rate equations are obtained by taking statistical expectation values of %%@
the EOMs, Eqs.~(\ref{eom:r00})-(\ref{eom:rdd}), which clearly involve the %%@
statistical averaging of products of one reservoir (phonon-bath) variable and %%@
one device variable, such as ${\cal I}=\langle \sum_{{\bf k}, \eta} V_{\eta 1} %%@
c_{\eta {\bf k}}^\dagger e_n^\dagger f_{1m} X_{1,nm}\rangle$ and ${\cal %%@
J}=\langle \sum_{p} \kappa_p (b_{p}^\dagger + b_{-p}) \sqrt{n+1} (e_n^\dagger %%@
e_{n+1}^\pdag - {\rm H.c.}) \rangle$.

To determine the products, ${\cal I}$ and ${\cal J}$, we proceed by deriving %%@
EOMs for the system, phonon-bath, and reservoir operators, %%@
$F_{1,nm}=e_n^\dagger f_{1m}^\pdag$, $\hat\rho_{00}^{n,n+1}=e_n^\dagger %%@
e_{n+1}^\pdag$, $b_{p}$, and $c_{\eta {\bf k}}$:
\begin{subequations}
\bn
i\dot{F}_{1,nm} &=& [e_n^\dagger f_{1m}, \widetilde{H}]_- = %%@
[\widetilde{\varepsilon}_{1} + (m-n)\omega_0] F_{1,nm} \cr
&& + [F_{1,nm}, \widetilde{H}_T]_- + [F_{1,nm}, \widetilde{H}_{vB}]_-, %%@
\label{ga} \\
i\dot{\hat\rho}_{00}^{n,n+1} &=& [e_n^\dagger e_{n+1}, \widetilde{H}]_- = %%@
\omega_0 \hat\rho_{00}^{n,n+1} + [\hat\rho_{00}^{n,n+1}, \widetilde{H}_T]_- %%@
\cr
&& + [\hat\rho_{00}^{n,n+1}, \widetilde{H}_{vB}]_-, \label{ga1} \\
i\dot{b}_{p} &=& [b_{p}, \widetilde{H}]_- = \omega_p b_p + [b_{p}, %%@
\widetilde{H}_{vB}]_-, \label{b} \\
i\dot c_{\eta {\bf k} } &=& [c_{\eta {\bf k} }, \widetilde H]_{-}= %%@
\epsilon_{\eta {\bf k} } c_{\eta {\bf k} } + [c_{\eta {\bf k} }, \widetilde %%@
H_T]_-. \label{a}    
\en
\end{subequations}
The EOM for $c_{\eta {\bf k} }^\dagger$ is easily obtained by Hermitian %%@
conjugation of the equations for $c_{\eta {\bf k} }$.
Formally integrating these equations, (\ref{ga})-(\ref{a}), from initial time %%@
$0$ to $t$ we obtain      
\begin{subequations}
\label{solution}
\bn
F_{1,nm}(t) &=& e^{-i [\widetilde{\varepsilon}_{1} + (m-n)\omega_0] t} %%@
F_{1,nm}(0)  \cr
&& \hspace{-1cm} - i \int_0^t dt' e^{-i [\widetilde{\varepsilon}_{1} + %%@
(m-n)\omega_0] \tau} [F_{1,nm}(t'), \widetilde H_T (t')]_- \cr
&& \hspace{-1cm} - i \int_0^t dt' e^{-i [\widetilde{\varepsilon}_{1} + %%@
(m-n)\omega_0] \tau} [F_{1,nm}(t'), \widetilde H_{vB} (t')]_-, \cr
&& \label{solution1} \\
\hat\rho_{00}^{n,n+1}(t) &=& e^{-i \omega_0 t} \hat\rho_{00}^{n,n+1}(0)  \cr
&& \hspace{-1cm} - i \int_0^t dt' e^{-i \omega_0 \tau} %%@
[\hat\rho_{00}^{n,n+1}(t'), \widetilde H_T (t')]_- \cr
&& \hspace{-1cm} - i \int_0^t dt' e^{-i \omega_0 \tau} %%@
[\hat\rho_{00}^{n,n+1}(t'), \widetilde H_{vB} (t')]_-, \cr
&& \label{solution1a} \\
b_{p}(t) &=& e^{-i \omega_p t} b_{p}(0) - i \int_0^t dt' e^{-i \omega_p \tau} %%@
\cr
&& \times [b_{p}(t'), \widetilde H_{vB}(t')]_-, \label{solution2} \\
c_{\eta {\bf k} }(t) &=& e^{-i\epsilon_{\eta {\bf k} } t} c_{\eta {\bf k} }(0) %%@
-i \int_0^t dt' e^{-i \epsilon_{\eta {\bf k} } \tau} \cr
&& \times [c_{\eta {\bf k} }(t'), \widetilde H_T(t')]_-, \label{solution3}
\en
\end{subequations}
with $\tau=t-t'$. In the absence of tunnel-coupling, $H_{I}\rightarrow 0$, we %%@
have
\begin{subequations}
\label{free}
\bn
F_{1,nm}^o(t)&=& e^{-i [\widetilde{\varepsilon}_{1} + (m-n)\omega_0] \tau} %%@
F_{1,nm}^o(t'), \label{free1} \\
\hat\rho_{00,o}^{n,n+1}(t)&=& e^{-i \omega_0 \tau} %%@
\hat\rho_{00,o}^{n,n+1}(t'), \label{free1a} \\
b_{p}^o(t)&=& e^{-i \omega_p \tau} b_{p}^o(t'), \label{free1a} \\
c_{\eta {\bf k} }^o(t) &=& e^{-i\epsilon_{\eta {\bf k} } \tau} c_{\eta {\bf k} %%@
}^o(t'). \label{free2} 
\en
\end{subequations} 

A standard assumption in the derivation of a quantum Langevin equation is that %%@
the time scale of decay processes is much slower than that of free evolution, %%@
which is reasonable in the weak-tunneling approximation. This bespeaks a %%@
dichotomy of time-developments of the involved operators into a %%@
rapidly-varying (free) part and a slowly-varying (dissipative) part. Focusing %%@
attention on the slowly-varying decay processes, and noting that the %%@
infinitude of macroscopic bath variables barely senses reaction from weak %%@
interaction with the QD, it is appropriate to substitute the time-dependent %%@
{\em decoupled} reservoir, phonon-bath, and QD operators of %%@
Eqs.~(\ref{free1})-(\ref{free2}) into the integrals on the right of %%@
Eqs.~(\ref{solution1})-(\ref{solution3}). This yields approximate results for %%@
the reservoir operators as:\cite{Schwinger,Ackerhalt,Dong2,Dong3}
\begin{subequations}
\label{coi}
\bq
c_{\eta {\bf k} }(t)= c_{\eta {\bf k} }^o(t)+ c_{\eta {\bf k} }^{rT}(t), %%@
\label{P}
\eq
with
\bq
c_{\eta {\bf k} }^{rT}(t)=-i \int_{0}^t d\tau [c_{\eta {\bf k} }^o(t), %%@
\widetilde H_{T}^o(t')]_{-}, \label{Pr}
\eq
\end{subequations}
where $\widetilde H_{T(vB)}^o$ is composed of the operators in $\widetilde %%@
H_{T(vB)}$ which are replaced by their decoupled counterparts (interaction %%@
picture). In fact, this is just the operator formulation of linear response %%@
theory.  
Similarly, the approximate results for the QD and phonon-bath are also divided %%@
into two parts:
\begin{subequations}
\bn
F_{1,nm}(t)&=& F_{1,nm}^{o}(t)+ F_{1,nm}^{rT}(t) + F_{1,nm}^{rvB}(t)\cr
&=& F_{1,nm}^{o}(t) -i \int_{0}^t d\tau [F_{1,nm}^{o}(t), \widetilde %%@
H_{T}^o(t')]_{-} \cr
&& -i \int_{0}^t d\tau [F_{1,nm}^{o}(t), \widetilde H_{vB}^o(t')]_{-}, %%@
\label{F} \\
\hat\rho_{00}^{n,n+1}(t)&=& \hat\rho_{00,o}^{n,n+1}(t)+ %%@
\hat\rho_{00,rT}^{n,n+1}(t) + \hat\rho_{00,rvB}^{n,n+1}(t)\cr
&=& \hat\rho_{00,o}^{n,n+1}(t) -i \int_{0}^t d\tau %%@
[\hat\rho_{00,o}^{n,n+1}(t), \widetilde H_{T}^o(t')]_{-} \cr
&& -i \int_{0}^t d\tau [\hat\rho_{00,o}^{n,n+1}(t), \widetilde %%@
H_{vB}^o(t')]_{-}, \label{F1} \\
b_{p}(t)&=& b_{p}^{o}(t)+ b_{p}^{rvB}(t) \cr
&=& b_{p}^{o}(t) -i \int_{0}^t d\tau [ b_{p}^{o}(t), \widetilde %%@
H_{vB}^o(t')]_{-}. \label{b1}
\en
\end{subequations}
Note that we use the super(sub)scripts $rT$ and $rvB$ denote the reactions %%@
from tunnel-coupling and environmental dissipation, respectively.  

Employing the approximate solutions of Eqs.~(\ref{P}), (\ref{Pr}) and %%@
(\ref{F}), we can evaluate ${\cal I}$ as
\bn
{\cal I} &=& \sum_{\eta, {\bf k}} V_{\eta 1} \langle [c_{\eta {\bf %%@
k}}^{o\dagger}(t) + c_{\eta {\bf k}}^{r\dagger}(t) ] \cr
&& \times [F_{1,nm}^o(t) + F_{1,nm}^r(t) ]X_{1,nm}\rangle \cr
&\simeq& \sum_{\eta, {\bf k}} V_{\eta 1} \langle [c_{\eta {\bf %%@
k}}^{o\dagger}(t) F_{1,nm}^r(t) + c_{\eta {\bf k}}^{r\dagger}(t) F_{1,nm}^o(t) %%@
] X_{1,nm}\rangle. \cr
&& 
\en
The statistical averages involved here can be taken separately in regard to %%@
the electron ensembles of the reservoirs (many degrees of freedom) and in %%@
regard to the few degrees of freedom of the QD EPDP states. Accordingly, the %%@
statistical average of the product of one reservoir operator and one system %%@
operator factorizes in the averaging procedure. Therefore, the statistical %%@
average of $c_{\eta {\bf k}}^{o\dagger}(t) F_{1,nm}^o(t)$ vanishes. Moreover, %%@
in the sequential picture of resonant tunneling, the tunneling rates and %%@
current are proportional to second-order tunnel-coupling matrix elements. We %%@
thus neglect the term $c_{\eta {\bf k}}^{r\dagger}(t) F_{1,nm}^r(t)$ as it is %%@
proportional to the third-order tunnel-coupling matrix element, $O(V_{\eta %%@
j}^3)$.   
The other interaction terms arise from tunneling reaction, and are of %%@
second-order of $V_{\eta j}$.
After some lengthy but straightforward algebraic calculations, we obtain
\begin{subequations}
\bn
{\cal I} &=& -i \sum_{\eta} \Gamma_{\eta 1} \int d\epsilon \int_0^t d\tau %%@
e^{i(\epsilon- \widetilde{\varepsilon}_1)\tau} \{ f_{\eta}(\epsilon) %%@
\rho_{00}^n \cr
&& - [1- f_{\eta}(\epsilon)] \rho_{11}^m \} \bar {X}_{1,nm}^\dagger %%@
X_{1,nm}^\pdag ,  \label{calI}
\en 
in which 
\bq
\Gamma_{\eta j}=2\pi \varrho_{\eta} |V_{\eta j}|^2
\eq
denotes the tunneling strength between the molecular orbital $j$ and lead %%@
$\eta$ ($\varrho_{\eta}$ is the density of states of lead $\eta$), %%@
$f_{\eta}(\epsilon)$ is the Fermi-distribution function of lead $\eta$ with %%@
temperature $T$, and
\bn
\bar X_{j,nm} &=& \langle n |X_{j}(t') |m\rangle,\\
\bar X_{j,nm}^\dagger &=& \langle m |X_{j}^\dagger(t') |n\rangle.
\en
Considering $a(t)=e^{-i\omega_0 \tau} a(t')$, we have
\bn
\bar X_{j,nm} &=& e^{i(m-n) \omega_0 \tau} X_{j,nm}, \\
\bar X_{j,nm}^\dagger &=& e^{-i(m-n) \omega_0 \tau} X_{j,nm}.
\en
\end{subequations}
In the derivation of Eq.~(\ref{calI}), we assume that states with different %%@
phonon-numbers are completely decoherent, $\rho_{00(dd)}^{nm}=\rho_{00(dd)}^n %%@
\delta_{nm}$ and $\rho_{jj}^{nm}=\rho_{jj}^n \delta_{nm}$, owing to big energy %%@
difference between the two direct product states $|j,n\rangle$ and %%@
$|j,m\rangle$ if $n\neq m$.  
Moreover, a Markovian approximation will be adopted by making the replacement
\bq
\int_{-\infty}^{t} d\tau \Longrightarrow \int_{-\infty}^{\infty} d\tau,
\eq
in the statistical averaging, since we are interested in the long time scale %%@
behavior of these density matrix elements.

Furthermore, ${\cal J}$ can be evaluated using Eqs.~(\ref{F1}), (\ref{b1}) %%@
with $b_p(t)=e^{-i\omega_p \tau} b_p(t')$ as
\bn
{\cal J} &\simeq& \sum_{p} \kappa_p \sqrt{n+1} \langle [(b_{p}^{o\dagger}(t) + %%@
b_{p}^o(t) ) \cr
&& \times (\rho_{00,rvB}^{n,n+1}(t) - \rho_{00,rvB}^{n+1,n}(t) ) \cr
&& + (b_{p}^{rvB\dagger}(t) + b_{p}^{rvB}(t) ) (\rho_{00,o}^{n,n+1}(t) - %%@
\rho_{00,o}^{n+1,n}(t)) ] \rangle \cr
&=& -i \sum_p \kappa_p^2 (n+1) \rho_{00}^n [ (n_B(\omega_p)+1) %%@
\delta(\omega_p+\omega_0) \cr
&& + n_B(\omega_p) \delta(\omega_p-\omega_0)] \cr
&& + i\sum_p \kappa_p^2 (n+1) \rho_{00}^{n+1} [ n_B(\omega_p) %%@
\delta(\omega_p+\omega_0) \cr
&& + (n_B(\omega_p)+1) \delta(\omega_p-\omega_0)],
\en
in which $n_B(\omega_p)=(e^{\omega_p/T}-1)^{-1}$ is the Bose-distribution %%@
function. With the assumption that the phonon-bath is composed of infinitely %%@
many harmonic oscillators having a wide and continuous spectral density, we %%@
can make the replacement, in the wide-band limit,
\bq
\sum_{p} \kappa_p^2 (\cdots) \longrightarrow  \int_{-\infty}^\infty d\omega_p %%@
\kappa_p^2 D(\omega_p) (\cdots),
\eq
yielding
\bq
{\cal J}= -i \varpi_p n_B(\omega_0) (n+1) \rho_{00}^n + i \varpi_p %%@
(n_B(\omega_0)+1) (n+1) \rho_{00}^{n+1},
\eq
with $\varpi_p=\kappa_0 \omega_0^\alpha$ being constant due to %%@
Eq.~(\ref{pbspectral}).  

Following the same calculational scheme indicated above, we evaluated the %%@
other statistical expectation values involved in the EOMs, %%@
Eqs.~(\ref{eom:r00})-(\ref{eom:rdd}). Finally, we obtained the following rate %%@
equations in terms of the direct product state representation of the %%@
density-matrix for the description of sequential tunneling through a molecular %%@
QD, accounting for the unequilibrated phonon effect and its modified tunneling %%@
rates, as well as IVM dissipation to the phonon environment:  
\begin{subequations}
\bn
\dot \rho_{00}^n &=& \sum_{m} \bigl [ \Gamma_{1,nm}^- \rho_{11}^m + %%@
\Gamma_{2,nm}^- \rho_{22}^m - (\Gamma_{1,nm}^+ + \Gamma_{2,nm}^+) \rho_{00}^n %%@
\bigr ] \cr
&& - (\varpi_{n}^+ +\varpi_{n}^-) \rho_{00}^n + \varpi_{n+1}^- \rho_{00}^{n+1} %%@
+ \varpi_{n-1}^+ \rho_{00}^{n-1}, \label{r00} \\
\dot \rho_{11}^n &=& \sum_{m} \bigl [\Gamma_{1,mn}^+ \rho_{00}^m - %%@
(\Gamma_{1,mn}^- + \widetilde {\Gamma}_{2,nm}^+) \rho_{11}^n + \widetilde %%@
{\Gamma}_{2,nm}^- \rho_{dd}^m \bigr ] \cr
&& - (\varpi_{n}^+ +\varpi_{n}^-) \rho_{11}^n + \varpi_{n+1}^- \rho_{11}^{n+1} %%@
+ \varpi_{n-1}^+ \rho_{11}^{n-1}, \\
\dot \rho_{22}^n &=& \sum_{m} \bigl [\Gamma_{2,mn}^+ \rho_{00}^m - %%@
(\Gamma_{2,mn}^- + \widetilde {\Gamma}_{1,nm}^+) \rho_{22}^n + \widetilde %%@
{\Gamma}_{1,nm}^- \rho_{dd}^m \bigr ] \cr
&& - (\varpi_{n}^+ +\varpi_{n}^-) \rho_{22}^n + \varpi_{n+1}^- \rho_{22}^{n+1} %%@
+ \varpi_{n-1}^+ \rho_{22}^{n-1}, \\
\dot \rho_{dd}^n &=& \sum_{m} \bigl [ \widetilde{\Gamma}_{2,mn}^+ \rho_{11}^m %%@
+ \widetilde{\Gamma}_{1,mn}^+ \rho_{22}^m -(\widetilde{\Gamma}_{1,mn}^- + %%@
\widetilde{\Gamma}_{2,mn}^-) \rho_{dd}^n \bigr ] \cr
&& - (\varpi_{n}^+ +\varpi_{n}^-) \rho_{dd}^n + \varpi_{n+1}^- \rho_{dd}^{n+1} %%@
+ \varpi_{n-1}^+ \rho_{dd}^{n-1}, \label{rdd}
\en
\end{subequations}
with the normalization relation $\sum_{n} (\rho_{00}^n +\rho_{11}^n %%@
+\rho_{22}^n + \rho_{dd}^n)=1$. The electronic tunneling rates are defined as
\begin{subequations}
\bn
\Gamma_{j,nm}^+ &=& \sum_{\eta} \Gamma_{\eta j,nm}^+= \sum_{\eta} \Gamma_{\eta %%@
j} \gamma_{nm}^j f_{\eta}(\widetilde {\epsilon}_j + (m-n)\omega_0), \cr
&& \\
\Gamma_{j,nm}^- &=& \sum_{\eta} \Gamma_{\eta j,nm}^- \cr
&=& \sum_{\eta} \Gamma_{\eta j} \gamma_{nm}^j [1-f_{\eta}(\widetilde %%@
{\epsilon}_j + (m-n)\omega_0)], \\
\widetilde {\Gamma}_{j,nm}^+ &=& \sum_{\eta} \widetilde {\Gamma}_{\eta j,nm}^+ %%@
\cr
&=& \sum_{\eta} \Gamma_{\eta j} \gamma_{nm}^j f_{\eta}(\widetilde U + %%@
\widetilde {\epsilon}_j + (m-n)\omega_0), \\
\widetilde {\Gamma}_{j,nm}^- &=& \sum_{\eta} \widetilde {\Gamma}_{\eta j,nm}^- %%@
= \sum_{\eta} \Gamma_{\eta j} \gamma_{nm}^j \cr
&& \times [1-f_{\eta}(\widetilde U + \widetilde{\epsilon}_j + (m-n)\omega_0)],
\en
with the {\em FC factor} ($p={\rm min}\{m,n\}$ and $q={\rm max}\{m,n\}$, %%@
denoting the smaller and larger of the quantities $m$ and $n$, respectively)
\bq
\gamma_{nm}^j = X_{j,nm}^2= e^{-g_j^2} g_j^{2|m-n|} \frac{p!}{q!}\bigl[ %%@
L_{p}^{|m-n|}(g_j^2)\bigr]^2,
\eq
\end{subequations}
describing the modification of tunnel-coupling due to phonon generation and %%@
emission during the electron tunneling processes. This FC factor is symmetric, %%@
$\gamma_{nm}^j=\gamma_{mn}^j$, and satisfies the sum rules $\sum_{n} %%@
\gamma_{nm}^j=\sum_{m} \gamma_{nm}^j=1$. Obviously, these rates have specific %%@
physical meanings: $\Gamma_{j,nm}^{+}$ ($\Gamma_{j,nm}^{-}$) describes the %%@
tunneling rate of an electron entering (leaving) level $j$ with null occupancy %%@
of level $\bar j$, together with the transition of vibrational quanta, %%@
$n\rightarrow m$ ($m\rightarrow n$); while $\widetilde\Gamma_{j,nm}^{+}$ %%@
($\widetilde\Gamma_{j,nm}^{-}$) describes the tunneling rate of an electron %%@
entering (leaving) level $j$ with level $\bar j$ occupied, together with the %%@
corresponding transition of the IVM state. 

The transition rates of phonon number states are
\begin{subequations}
\bn
\varpi_{n}^+ &=& \varpi_p n_B(\omega_0) (n+1), \\
\varpi_{n}^- &=& \varpi_p (n_B(\omega_0)+1)n,
\en   
\end{subequations}
which indicates that the state of the IVM changes from $n$ to $n+1$ ($n-1$) by %%@
absorbing (emitting) a phonon from (to) the phonon-bath without change of the %%@
electronic state. Note that $\varpi_n^++ \varpi_n^-$ defines the relaxation %%@
rate of the number state $n$ of the IVM due to dissipative coupling to the %%@
environment. Moreover, it should be pointed out that $\varpi_p=0$ denotes no %%@
dissipation of the IVM to the environment, signifying the maximal %%@
unequilibrated phonon effect in tunneling; while $\varpi_p=\infty$ denotes %%@
extremely strong dissipation of the IVM, so that it is always functions as an %%@
equilibrated state during each tunneling process, i.e., the excited phonon %%@
relaxes very quickly due to strong dissipation, before the next electronic %%@
tunneling event takes place. Obviously, strong dissipation, $\varpi_p=\infty$, %%@
forces the probability distributions on the right-hand side of %%@
Eqs.~(\ref{r00})-(\ref{rdd}) to have the forms, $\rho_{00}^n=\rho_{00}P^n$, %%@
$\rho_{jj}^n=\rho_{jj}P^n$, and $\rho_{dd}^n=\rho_{dd}P^n$, with a thermal %%@
phonon distribution $P^n=e^{-n\omega_0/T}(1-e^{-\omega_0/T})$ (assuming the %%@
phonon bath to have the same temperature as the electrodes).     

The tunneling current operator through the molecular QD is defined as the time %%@
rate of change of the charge density, $N_{\eta}=\sum_{{\bf k}} a_{\eta {\bf k} %%@
}^{\dagger} a_{\eta {\bf k} }^{\pdag}$, in lead $\eta$:
\bq
J_{\eta}= - \dot N_{\eta} = i [N_{\eta}, \widetilde H]_-= i [N_{\eta}, %%@
\widetilde H_I]_- . \label{i}
\eq
Employing linear-response theory in the interaction picture,\cite{Mahan} we %%@
have
\bn
I = \langle J_{L} \rangle =-i \int_{-\infty}^t dt' \langle [ J_{L}(t), %%@
\widetilde H_{I}^o(t')]_- \rangle .
\en  
Following the same procedures indicated above,\cite{Dong2,Dong3} we obtain the %%@
current formula through the left lead in terms of the density matrix elements %%@
of direct product states as
\bn
I &=& \sum_{nm} \bigl[ (\Gamma_{L1,nm}^+ + \Gamma_{L2,nm}^+) \rho_{00}^n + %%@
(\widetilde {\Gamma}_{L2,nm}^+ - \Gamma_{L1,mn}^-) \rho_{11}^n \cr
&& + (\widetilde {\Gamma}_{L1,nm}^+ - \Gamma_{L2,mn}^-) \rho_{22}^n - %%@
(\widetilde {\Gamma}_{L1,mn}^- + \widetilde {\Gamma}_{L2,mn}^-) \rho_{dd}^n %%@
\bigr]. \cr
&& \label{currentL}
\en

\section{MacDonald's formula for zero-frequency shot noise}

In this section, we discuss the zero-frequency current noise of %%@
vibration-mediated sequential tunneling through a molecular QD involving an %%@
unequilibrated phonon. For this purpose, we employ MacDonald's formula for %%@
shot noise\cite{MacDonald} based on a number-resolved version of the rate %%@
equations describing the number of completed tunneling events.\cite{Chen} This %%@
can be derived straightforwardly from the established QREs, %%@
Eqs.~(\ref{r00})--(\ref{rdd}). We introduce the two-terminal number-resolved %%@
density matrices $\rho_{jj}^{n(l,l')}(t)$, representing the probability that %%@
the system is in the electronic state $|j\rangle$ ($j=\{0,1,2,d\}$) with $n$ %%@
vibrational quanta at time $t$ together with $l(l')$ electrons occupying the %%@
left(right) lead due to tunneling events. Obviously, %%@
$\rho_{jj}^{n}(t)=\sum_{l,l'} \rho_{jj}^{n(l,l')}(t)$ and the resulting %%@
two-terminal number-resolved QREs for the case of an unequilibrated phonon %%@
are:
\begin{subequations}
\bn
\dot \rho_{00}^{n(l,l')} &=& \sum_{m} \bigl [ \Gamma_{L1,nm}^- %%@
\rho_{11}^{m(l-1,l')} + \Gamma_{L2,nm}^- \rho_{22}^{m(l-1,l')} \cr
&& + \Gamma_{R1,nm}^- \rho_{11}^{m(l,l'-1)} + \Gamma_{R2,nm}^- %%@
\rho_{22}^{m(l,l'-1)} \cr
&& - (\Gamma_{1,nm}^+ + \Gamma_{2,nm}^+) \rho_{00}^{n(l,l')} \bigr ] - %%@
(\varpi_{n}^+ +\varpi_{n}^-) \rho_{00}^{n(l,l')} \cr
&& + \varpi_{n+1}^- \rho_{00}^{n+1(l,l')} + \varpi_{n-1}^+ %%@
\rho_{00}^{n-1(l,l')}, \label{ttqre:r00} \\
\dot \rho_{11}^{n(l,l')} &=& \sum_{m} \bigl [\Gamma_{L1,mn}^+ %%@
\rho_{00}^{m(l+1,l')} + \Gamma_{R1,mn}^+ \rho_{00}^{m(l,l'+1)} \cr
&& - (\Gamma_{1,mn}^- + \widetilde {\Gamma}_{2,nm}^+) \rho_{11}^{n(l,l')} + %%@
\widetilde {\Gamma}_{L2,nm}^- \rho_{dd}^{m(l-1,l')} \cr
&& + \widetilde {\Gamma}_{R2,nm}^- \rho_{dd}^{m(l,l'-1)} \bigr ] - %%@
(\varpi_{n}^+ +\varpi_{n}^-) \rho_{11}^{n(l,l')} \cr
&& + \varpi_{n+1}^- \rho_{11}^{n+1(l,l')} + \varpi_{n-1}^+ %%@
\rho_{11}^{n-1(l,l')}, \\
\dot \rho_{22}^{n(l,l')} &=& \sum_{m} \bigl [\Gamma_{L2,mn}^+ %%@
\rho_{00}^{m(l+1,l')} + \Gamma_{R2,mn}^+ \rho_{00}^{m(l,l'+1)} \cr
&& - (\Gamma_{2,mn}^- + \widetilde {\Gamma}_{1,nm}^+) \rho_{22}^{n(l,l')} + %%@
\widetilde {\Gamma}_{L1,nm}^- \rho_{dd}^{m(l-1,l')} \cr
&& + \widetilde {\Gamma}_{R1,nm}^- \rho_{dd}^{m(l,l'-1)} \bigr ] - %%@
(\varpi_{n}^+ +\varpi_{n}^-) \rho_{22}^{n(l,l')} \cr
&& + \varpi_{n+1}^- \rho_{22}^{n+1(l,l')} + \varpi_{n-1}^+ %%@
\rho_{22}^{n-1(l,l')}, \\
\dot \rho_{dd}^{n(l,l')} &=& \sum_{m} \bigl [ \widetilde{\Gamma}_{L2,mn}^+ %%@
\rho_{11}^{m(l+1,l')} + \widetilde{\Gamma}_{R2,mn}^+ \rho_{11}^{m(l,l'+1)} \cr
&& + \widetilde{\Gamma}_{L1,mn}^+ \rho_{22}^{m(l+1,l')} + %%@
\widetilde{\Gamma}_{R1,mn}^+ \rho_{22}^{m(l,l'+1)} \cr && %%@
-(\widetilde{\Gamma}_{1,mn}^- + \widetilde{\Gamma}_{2,mn}^-) %%@
\rho_{dd}^{n(l,l')} \bigr ]
- (\varpi_{n}^+ +\varpi_{n}^-) \rho_{dd}^{n(l,l')} \cr
&& + \varpi_{n+1}^- \rho_{dd}^{n+1(l,l')} + \varpi_{n-1}^+ %%@
\rho_{dd}^{n-1(l,l')}. \label{ttqre:rdd}
\en
\end{subequations}

The current flowing through the system can be evaluated by the time rates of %%@
change of electron numbers in lead $\eta$ as
\bq
I_{\eta} = \dot N_{\eta}(t) =\frac{d}{dt} \sum_{l,l'} l_{\eta} P^{(l,l')}(t) %%@
{\Big |}_{t\rightarrow\infty}, \label{Inr}
\eq
where
\bq
P^{(l,l')}(t) = \sum_{n} [\rho_{00}^{n(l,l')}(t)+ \rho_{11}^{n(l,l')}(t)+ %%@
\rho_{22}^{n(l,l')}(t) + \rho_{dd}^{n(l,l')}(t) ]
\eq
is the total probability of transferring $l(l')$ electrons into the %%@
left(right) lead by time $t$ and $l_{\eta}=l(l')$ with $\eta=L(R)$. It is %%@
readily verified that the current obtained from Eq.~(\ref{Inr}) by means of %%@
the number-resolved QREs, Eqs.~(\ref{ttqre:r00})--(\ref{ttqre:rdd}), is %%@
exactly the same as that obtained from Eq.~(\ref{currentL}). The %%@
zero-frequency shot noise with respect to lead $\eta$ is similarly defined in %%@
terms of $P^{(l,l')}(t)$ as well:\cite{Dong3,MacDonald,Chen}
\bq
S_{\eta}(0)=2\frac{d}{dt} \left [ \sum_{l,l'} l_{\eta}^2 P^{(l,l')}(t) - (t %%@
I_{\eta})^2 \right ] {\Big |}_{t\rightarrow\infty}. \label{snnr}
\eq

To evaluate $S_{\eta}(0)$, we define an auxiliary function $G_{jj}^{\eta %%@
n}(t)$ as
\bq
G_{jj}^{\eta n}(t) = \sum_{l,l'} l_{\eta} \rho_{jj}^{n(l,l')}(t),
\eq
whose equations of motion can be readily deduced employing the number-resolved %%@
QREs, Eqs~(\ref{ttqre:r00})--(\ref{ttqre:rdd}), in matrix form: $\dot{\bm %%@
G}^{\eta}(t)={\cal M}_{\eta} {\bm G}^{\eta }(t) + {\cal G}_{\eta} {\bm %%@
\rho}(t)$ with ${\bm G}^{\eta }(t)=({\bm G}_{00}^{\eta }, {\bm G}_{11}^{\eta}, %%@
{\bm G}_{22}^{\eta}, {\bm G}_{dd}^{\eta })^{T}$ and ${\bm \rho}(t)=({\bm %%@
\rho}_{00}, {\bm \rho}_{11}, {\bm \rho}_{22}, {\bm \rho}_{dd})^{T}$ [here %%@
${\bm G}_{jj}^\eta=(G_{jj}^{\eta 0}, G_{jj}^{\eta 1}, \cdots)^T$ and ${\bm %%@
\rho}_{jj}=(\rho_{jj}^0, \rho_{jj}^1, \cdots)^T$]. ${\cal M}_{\eta}$ and %%@
${\cal G}_{\eta}$ can be obtained easily from %%@
Eqs.~(\ref{ttqre:r00})--(\ref{ttqre:rdd}). 
Applying the Laplace transform to these equations yields
\bq
{\bm G}^{\eta}(s) = (s {\bm I}-{\cal M}_{\eta})^{-1} {\cal G}_{\eta} {\bm %%@
\rho}(s),
\eq
where ${\bm \rho}(s)$ is readily obtained by applying the Laplace transform to %%@
its equations of motion with the initial condition ${\bm \rho}(0)={\bm %%@
\rho}_{st}$ [${\bm \rho}_{st}$ denotes the stationary solution of the QREs, %%@
Eqs~(\ref{r00})--(\ref{rdd})]. Due to the inherent long-time stability of the %%@
physical system under consideration, all real parts of nonzero poles of ${\bm %%@
\rho}(s)$ and ${\bm G}^{\eta}(s)$ are negative definite. Consequently,  the %%@
divergent terms arising in the partial fraction expansions of ${\bm \rho}(s)$ %%@
and ${\bm G}^{\eta}(s)$ as $s\rightarrow 0$ entirely determine the large-$t$ %%@
behavior of the auxiliary functions, i.e. the zero-frequency shot noise, %%@
Eq.~(\ref{snnr}). 

It is worth noting that (1) our two-terminal number-resolved QREs, %%@
Eqs.~(\ref{ttqre:r00})--(\ref{ttqre:rdd}), facilitate evaluation of the %%@
bias-voltage dependent zero-frequency shot noise for arbitrary interdot %%@
hopping; (2) our calculations yield $S_{L}(0)=S_{R}(0)$.

\section{Results and discussion}

We now proceed with numerical calculations of the current $I$ %%@
[Eq.~(\ref{currentL})], the zero-frequency current noise $S(0)$ and the Fano %%@
factor $F=S(0)/2I$ for the two-MO model in order to explain the particular %%@
experimental data in the current-voltage ($I$-$V$) characteristics recently %%@
reported for a suspended CNT.\cite{Sapmaz} 

For this purpose, we set the parameters in our calculations as: $\omega_0=1$ %%@
as the energy unit, $\Gamma_{R1}/\Gamma_{L1}=10^3$, %%@
$\Gamma_{L2}/\Gamma_{L1}=1$, $\Gamma_{R2}/\Gamma_{L1}=0.1$, $g_1=1$, and %%@
$\widetilde{\varepsilon}_1=0$, $\widetilde{\varepsilon}_2=4.0\omega_0$. For %%@
simplicity, we fix the energy of the ground MO to be zero and ignore the %%@
nonzero bias-voltage-induced energy shift of the MO, which can be achieved %%@
using gate voltage in the experiment. Different from Ref.~\onlinecite{Nowack}, %%@
we choose a large asymmetry in the tunneling rates of the first MO, %%@
$\Gamma_{R1}/\Gamma_{L1}=10^3$, and an intermediate electron-phonon coupling %%@
strength, $g_1$. In particular, numerical fits of the experimentally measured %%@
data for the $I$-$V$ curves show $g_1\simeq 1$ for long CNTs.\cite{Sapmaz} %%@
These two parameters are believed to be necessary for the appearance of NDC in %%@
combination with the unequilibrated phonon condition, for the regime in which %%@
the ground MO is predominant in tunneling (this will be shown %%@
below).\cite{Zazunov} Moreover, we also choose a large asymmetry in the %%@
tunneling rates of the second MO, $\Gamma_{R2}/\Gamma_{L2}=0.1$, which is %%@
responsible for the appearance of strong NDC when the excited MO is %%@
predominant in tunneling as pointed out in Ref.~\onlinecite{Hettler}. %%@
Furthermore, we examine the effect of EPC in the excited MO on the strong NDC. 

In our calculations, $\varpi_p=0$ denotes no dissipation of the IVM to the %%@
environment, corresponding to the maximal uneqilibrated phonon effect in %%@
resonant tunneling; increasing dissipation strength, $\varpi_p>0$,  describes %%@
the action of the dissipative environment as it begins to relax the excited %%@
IVM towards an equilibrium phonon state (in the limit $\varpi_p=\infty$). This %%@
parameter allows one to examine the continuous variation of the effect of the %%@
dissipative environment on the excited IVM, which is helpful in developing a %%@
deep understanding of the underlying properties of vibration-mediated resonant %%@
tunneling and its fluctuations in CNTs.\cite{Haupt} Throughout the paper we %%@
set the temperature as $T=0.02\omega_0$ and assume that the bias voltage $V$ %%@
is applied symmetrically, $\mu_L=-\mu_R=V/2$.

\subsection{{\em Strong NDC} in current-voltage characteristics}

Figure 1 exhibits the calculated results without environmental dissipation, %%@
$\varpi_p=0$, for the $I$-$V$ characteristic and the occupation probability of %%@
the excited MO, $\rho_{22}$, in the case of strong Coulomb blockade %%@
$\widetilde{U}=\infty$. If the excited MO is weakly coupled to the IVM, for %%@
instance $g_2=0.01$ (solid lines) in Fig.~1, we find that (1) the ground MO %%@
plays a dominant role in tunneling if $V\leq 2 %%@
\widetilde{\varepsilon}_2$($\simeq 8$); (2) a series of equally spaced steps %%@
in voltage appears in the $I$-$V$ curve demonstrating vibration-mediated %%@
transport behavior, associated with a gradually reduced step height due to the %%@
{\em FC factor} in this voltage %%@
range;\cite{Bose,Alexandrov,McCarthy,Mitra,Koch} (3) a {\em weak NDC} and %%@
current peak occur at the onset of each step beginning from the second step, %%@
which are ascribed to the combination of asymmetric geometry, %%@
$\Gamma_{R1}/\Gamma_{L1}=10^3$,\cite{Zazunov} and the unequilibrated phonon %%@
effect (this will be shown below); and (4) a {\em strong NDC} suddenly emerges %%@
when the excited MO becomes active in tunneling at %%@
$V=2\widetilde{\varepsilon}_2$, which helps to explain the experimental data %%@
[Fig.~3(a) in Ref.~\onlinecite{Sapmaz}]. As pointed out previously by M. %%@
Hettler {\it et al.}, who studied nonlinear transport through a molecular QD %%@
without IVM,\cite{Hettler} this current decrease is a combined effect of %%@
strong Coulomb blockade and weak tunnel-coupling of the excited MO to the %%@
electrodes. It is for this reason that we selected %%@
$\Gamma_{R2}/\Gamma_{L1}=0.1$ in our numerical calculations. When the bias %%@
voltage increases to activate the excited MO, an electron can occupy this MO %%@
but does not easily tunnel out due to the weaker escape rate $\Gamma_{R2}$, as %%@
shown by the solid line in Fig.~1(b), thus blocking occupation of the ground %%@
MO because of the strong Coulomb repulsion. As a result, the tunneling current %%@
depends mainly on the contribution of the excited MO, leading to a suppression %%@
of current magnitude by about a factor $\Gamma_{R2}^2$. Relaxing either of %%@
these two conditions results in the elimination of the strong NDC. For %%@
example, the inset of Fig.~1 plots the corresponding results for the case %%@
without Coulomb interaction. Figure 2 shows the $I$-$V$ curve with increasing %%@
$\Gamma_{R2}$.                

\begin{figure}[htb]
\includegraphics[height=7cm,width=8cm]{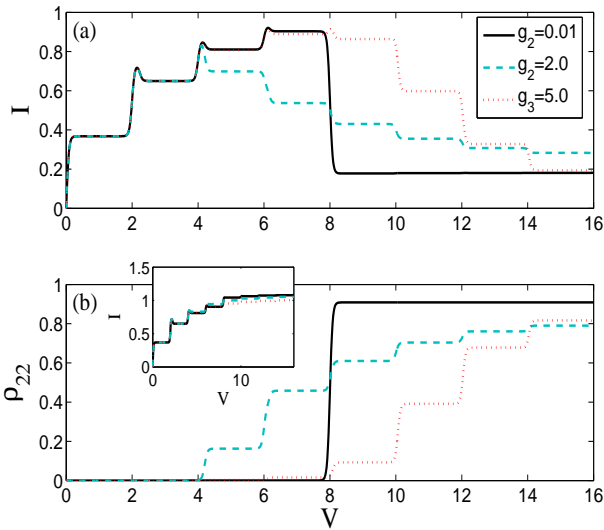}
\caption{(a) Calculated current $I$, and (b) occupation probability %%@
$\rho_{22}$ as functions of bias voltage in the case of strong Coulomb %%@
blockade, $\widetilde{U}=\infty$, and unequilibrated phonon, $\varpi=0$ for %%@
$g_2=0.01$, $2.0$, and $5.0$. Inset: current $I$ vs. voltage $V$ in the case %%@
$\widetilde{U}=0$.}
\label{fig1}
\end{figure}

\begin{figure}[htb]
\includegraphics[height=3.5cm,width=8cm]{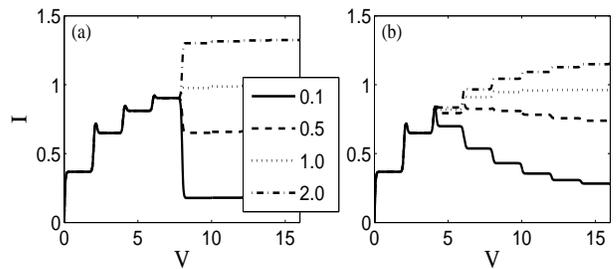}
\caption{Current-voltage characteristics for several values of increasing %%@
tunneling rate $\Gamma_{R2}/\Gamma_{L1}$ with (a) $g_2=0.01$ and (b) $g_2=2.0$ %%@
in the case of strong Coulomb blockade, $\widetilde{U}=\infty$, and %%@
unequilibrated phonon, $\varpi=0$.}
\label{fig2}
\end{figure}

Interestingly, if the excited MO is also appreciably coupled to the IVM, a %%@
stepdown behavior with equally spaced width in voltage is superposed on the %%@
overall decrease of current, leading to a slowdown of the original %%@
rapid-reduction of current and more peaks of the NDC. More interestingly, an %%@
advancing and/or a postponing of the current decrease are observed, depending %%@
upon the relative strengths of the EPCs of the ground and excited MOs. %%@
Intuitively, this advancing and postponing can be ascribed to the %%@
corresponding behaviors of the occupation probability of the second MO in the %%@
presence of EPC, as shown in Fig.~1(b). In the following, we will offer a %%@
deeper theoretical explanation of these results.      

\begin{figure}[htb]
\includegraphics[height=7cm,width=8.5cm]{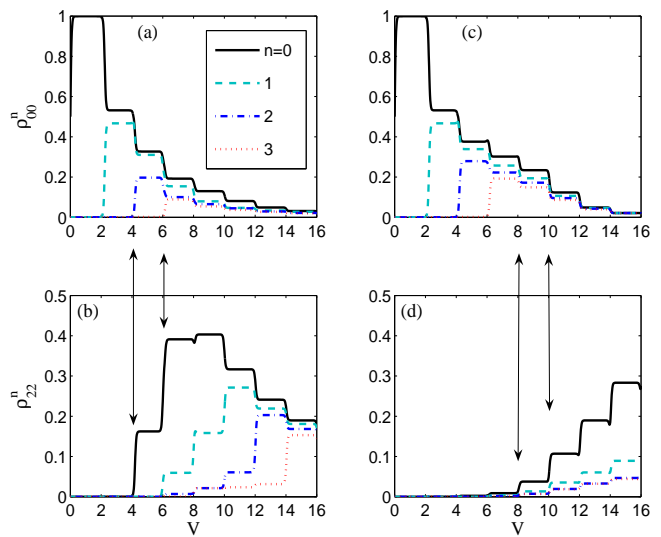}
\caption{Electron-phonon joint occupation probabilities, $\rho_{00}^n$ (a,c) %%@
and $\rho_{22}^n$ (b,d), vs. bias voltage relevant to Fig.~1 parameters; for %%@
(a,b) $g_2=2.0$ and for (c,d) $g_2=5.0$.}
\label{fig3}
\end{figure}

\begin{figure}[htb]
\includegraphics[height=3.5cm,width=5.5cm]{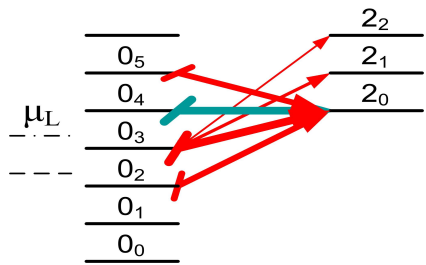}
\caption{Allowed transitions for Fig.~3 (see text for details).}
\label{fig4}
\end{figure}

For illustrative purposes, it is helpful to examine bias voltage dependences %%@
of the electron-phonon joint occupation probabilities (EPJOPs), $\rho_{00}^n$ %%@
and $\rho_{22}^n$. We plot the calculated results in Fig.~3 for two cases with %%@
$g_2=2.0$ (a,b) and $g_2=5.0$ (c,d). It is unnecessary to plot the EPJOPs for %%@
the ground MO, $\rho_{11}^n$, because once an electron enters into the first %%@
MO from the left electrode, it will escape very rapidly to the right electrode %%@
due to the strongly asymmetric configuration, $\Gamma_{R1}/\Gamma_{L1}=10^3$ %%@
(We only consider $V>0$ in this paper). Therefore, all $\rho_{11}^n$ are %%@
nearly zero. In contrast to this, once an electron is injected into the second %%@
MO, it is effectively trapped in this MO due to the suppressed tunnel-out rate %%@
$\Gamma_{R2}/\Gamma_{L1}=0.1$ in the present model under investigation. With %%@
regard to this consideration, we first discuss the results for $g_2=2.0$. %%@
Fig.~3(a) clearly shows that the EPDP state $|0\rangle_1 |0\rangle_2 \otimes %%@
|n\rangle$ (for notational convenience, we use $0_n$ to denote this EPDP state %%@
and $2_n$ to represent $|0\rangle_1 |1\rangle_2 \otimes |n\rangle$ below) is %%@
occupied, and contributes to current when the Fermi energy of the left lead is %%@
equal to the energy of the corresponding direct product state, $\mu_L= n %%@
\omega_0$, as illustrated in the schematic energy diagram, Fig.~4. Opening a %%@
new channel will cause a decrease of the occupation probabilities of previous %%@
direct product states. Intuitively, one might think that the EPDP state %%@
$|0\rangle_1 |1\rangle_2 \otimes |0\rangle$($\equiv 2_0$) is unoccupied until %%@
the bias voltage increases to $V=2\widetilde{\varepsilon}_2=8.0\omega_0$. From %%@
Fig.~3(b), however we (surprisingly) observe that the EPDP state $2_0$ is %%@
actually becoming occupied even at $V=4.0\omega_0$, which is half of the %%@
conventional resonant tunneling value. Moreover, the EPDP states $2_1$ and %%@
$2_2$ are both becoming occupied starting at $V=6.0\omega_0$, albeit their %%@
corresponding resonant values should traditionally be $V=10.0\omega_0$ and %%@
$12.0\omega_0$ respectively. In addition, differing from the voltage %%@
dependence features of $\rho_{00}^n$, the opening of new channels involving %%@
the excited MO does not cause a reduction of the occupation probabilities of %%@
previous channels. These peculiar properties can be understood qualitatively %%@
in terms of phonon-induced {\em cascaded} single-electron transitions as %%@
illustrated in Fig.~4. As pointed out by M.R. Wegewijs, {\it et al}., %%@
arbitrarily high vibrational excitations can in principle be accessed via {\em %%@
cascades} of single-electron tunneling processes driven by a finite bias %%@
voltage, since transitions between the EPDP states is related to the variation %%@
of electronic energy and the {\em change} of phonon-number states. For %%@
instance, if the Fermi energy of the left lead is located between the EPDP %%@
states $0_2$ and $0_3$ with increasing bias voltage, i.e., %%@
$4.0\omega_0<V<6.0\omega_0$, the EPDP state $0_2$ becomes occupied as shown in %%@
Fig.~3(a). Moreover, the single-electron transition, $0_2\rightarrow 2_0$, %%@
indicated by the arrow in Fig.~4 is {\em also} permitted because the bias %%@
voltage $V=4.0\omega_0$ provides sufficient energy to activate this %%@
transition, $\widetilde{\varepsilon}_2+(0-2)\omega_0=2.0\omega_0$. As a %%@
result, albeit the Fermi energy of the left lead is not aligned with the %%@
energy $\widetilde{\varepsilon}_2=4.0\omega_0$ of the EPDP state $2_0$, this %%@
state also becomes occupied [Fig.~3(b)], which precedes the conventional %%@
resonance value $V=8.0\omega_0$. Likewise, when the bias voltage increases to %%@
$V>6.0\omega_0$, the EPDP state $0_3$ becomes occupied, and concomitantly, the %%@
transitions $0_3\rightarrow 2_0$, $0_3\rightarrow 2_1$, $0_3\rightarrow 2_2$, %%@
and even $0_1\rightarrow 2_0$ are also permitted with differing transition %%@
rates depending on the FC factors (denoted by the different widths of the %%@
arrows in Fig.~4) {\em via the cascade transition mechanism}. Therefore, we %%@
observe from Fig.~3(b) that the states $2_1$ and $2_2$ both become occupied %%@
starting at $V=6.0\omega_0$. In contrast to the situation in %%@
Ref.~\onlinecite{Nowack}, the back-transition $2_n\rightarrow 0_m$ is %%@
prohibited in the present model due to the above-mentioned trapping effect of %%@
the excited MO (stemming from the suppressed escape rate $\Gamma_{R2}$). %%@
Consequently, we find an accumulated increase of occupation probabilities of %%@
the EPDP states with low vibrational excitations up to a threshold value of %%@
bias voltage, in which a considerably large number of channels are stimulated %%@
and become active. In sum, the EPC-induced unidirectional cascaded transitions %%@
are responsible for the advanced appearance of NDC at lower bias voltages than %%@
one might otherwise expect.

We now turn to examine the mechanism of the postponed appearance of NDC in the %%@
case of $g_2=5.0\gg g_1$. Figures 3(c) and (d) show that the occupation %%@
probabilities of $\rho_{00}^n$ have a similar bias voltage dependence to that %%@
of the case of $g_2=2.0\sim g_1$, but the situation is considerably different %%@
for $\rho_{22}^n$: obviously, $\rho_{22}^0$ is nearly zero until the bias %%@
voltage increases up to $10.0\omega_0$, which is even higher than the %%@
resonance value, $V=2\widetilde{\varepsilon}_2=8.0\omega_0$. Therefore, it is %%@
interesting to explore why the EPC-induced cascade mechanism of %%@
single-electron transitions does not work in this situation. Albeit the %%@
cascade mechanism for the transition, e.g. $0_2\rightarrow 2_0$, satisfies the %%@
usual resonance condition from the energetic point of view, the actual %%@
occurrence of this transition still depends on the relevant transition rate %%@
determined by the corresponding FC factor, $\gamma_{02}^2$. In Fig.~5, we show %%@
the FC factors $\gamma_{nm}^2$ from $n,m=0$ to $10$ for $g_2=2.0$ and $5.0$, %%@
respectively. Clearly, a nearly vanishing FC factor $\gamma_{02}^2$ occurs if %%@
$g_2=5.0$. The strong EPC strength even blocks the conventional resonant %%@
transition $0_4\rightarrow 2_0$. Only when a large number of transitions are %%@
opened by a sufficiently high bias voltage (e.g. $V=10.0\omega_0$ here), can a %%@
significant occupation of $\rho_{22}$ be accumulated, thus leading to the %%@
current decrease.          

\begin{figure}[htb]
\includegraphics[height=3.5cm,width=8.5cm]{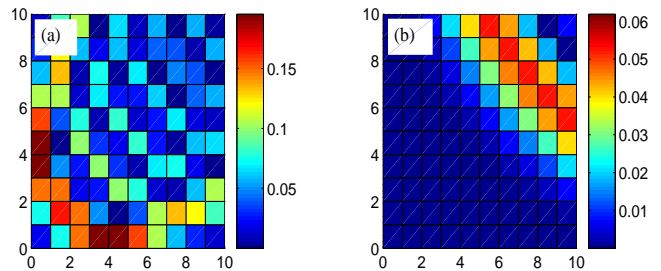}
\caption{The FC factors, $\gamma_{nm}^{2}$ ($n,m=0$--$10$), for (a) $g_2=2.0$ %%@
and (b) $g_2=5.0$.}
\label{fig5}
\end{figure}

We now further examine the disappearance of NDC shown in Fig.~2(b). With %%@
gradual relaxation of the trapping effect of the excited MO (increasing %%@
$\Gamma_{R2}$), the unidirectional cascaded transitions becomes bidirectional, %%@
i.e., not only can the transitions $0_n\rightarrow 2_m$ occur, but also %%@
$2_n\rightarrow 0_m$ occurs as well, provided that the energy conservation %%@
condition is satisfied and the FC factor permits. Therefore, the inverse %%@
transition, $2_0\rightarrow 0_2$, reduces the occupation of the state $2_0$ %%@
(not shown here), and finally reduces the NDC.

\subsection{Effects of dissipation to environment and super-Poissonian current %%@
noise}

We now discuss the effects of environmental dissipation on the current and %%@
zero-frequency shot noise. Figures 6 and 7 exhibit the dissipation dependences %%@
of the current $I$ and Fano factor $F=S(0)/2I$ as functions of bias voltage %%@
for systems having $g_2=0.01$ and $2.0$, respectively. 

\begin{figure}[htb]
\includegraphics[height=6cm,width=8cm]{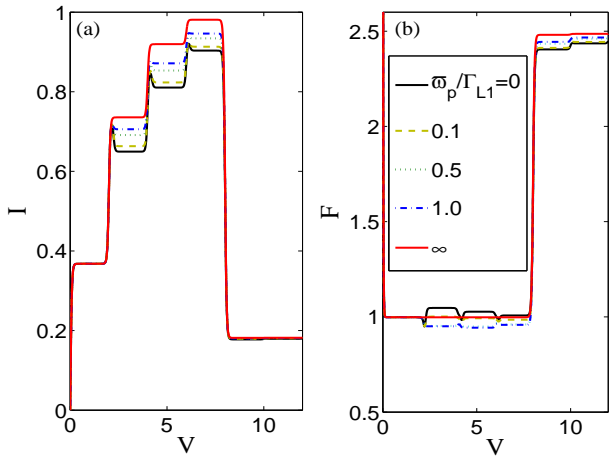}
\caption{Calculated current, $I$ (a) and Fano factor, $F=S(0)/2I$ (b) vs. bias %%@
voltage for the case of $g_2=0.01$ and various environmental dissipation rates %%@
$\varpi_p$. Other parameters are the same as in Fig.~1.}
\label{fig6}
\end{figure}

\begin{figure}[htb]
\includegraphics[height=6cm,width=8cm]{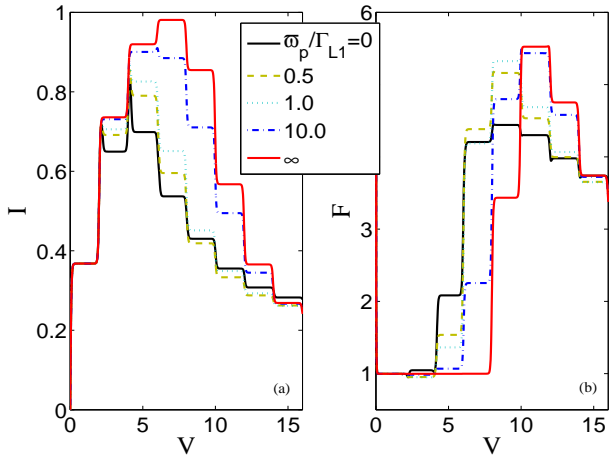}
\caption{The same figure as Fig.~6 except that $g_2=2.0$.}
\label{fig7}
\end{figure}

In the region where the ground MO is dominant in tunneling, it is clearly %%@
observed that the weak NDC becomes ever weaker with a gradually increasing %%@
dissipation rate $\varpi$. The $I$-$V$ characteristic finally exhibits only %%@
positive differential conductance at $\varpi=\infty$, i.e., the equilibrated %%@
phonon condition. Therefore, one can conclude that the observation of weak NDC %%@
in the recent transport measurements clearly indicates that the external %%@
voltage-driven unequilibrated phonon effect in a suspended CNT plays an %%@
essential role in determining its underlying transport %%@
properties.\cite{Sapmaz} Moreover, it is also clear that, due to the %%@
unequilibrated phonon effect, the current noise shows a weak super-Poissonian %%@
characteristic ($F>1$) associated closely with the appearance of NDC. %%@
Furthermore, the environmental dissipation suppresses the Fano factor and for %%@
certain values of the dissipation rate, $\varpi$, the shot noise shows weak %%@
sub-Poissonian behavior ($F<1$), but it becomes Poissonian ($F=1$) at the %%@
equilibrated phonon condition. This is just the traditional value of the Fano %%@
factor, $F=(\Gamma_{L1}^2 + \Gamma_{R1}^2)/(\Gamma_{L1} + \Gamma_{R1})^2$, for %%@
the extremely asymmetrical configuration $\Gamma_{R1}/\Gamma_{L1}=10^3$.  

From Figs.~6 and 7, we find strong super-Poissonian shot noise as a companion %%@
to the strong NDC when the second MO starts to contribute to the current, %%@
irrespective of the environmental dissipation. When the second MO is also %%@
coupled to the IVM, the environmental dissipation strongly influences the %%@
current and noise as shown in Fig.~7. In particular, the threshold value of %%@
bias voltage for strong NDC and strong enhancement of shot noise is exactly %%@
the traditional resonant tunneling value of the second MO.

\section{Conclusions}   

In summary, we have fully analyzed the external-bias-voltage-driven %%@
nonequilibrated phonon effect on nonlinear tunneling through a suspended CNT %%@
in the sequential tunneling regime. In order to qualitatively address the %%@
recent experimental results, we have modeled the CNT as a molecular QD having %%@
two electronic MOs with asymmetric tunnel-coupling rates to the left and right %%@
electrodes and strong interaction with an IVM. To study the role of %%@
dissipation of unequilibrated phonons, we have assumed further that the %%@
molecular IVM is also weakly coupled to a phonon bath ``environment". 
To carry out this analysis, we established generic rate equations in terms of %%@
the EPDP state and auxiliary-particle representation for the description of %%@
vibration-mediated resonant tunneling employing a microscopic quantum Langevin %%@
equation approach in the limit of weak tunneling and weak dissipation.  

Employing the ensuing rate equations derived here, we systematically analyzed %%@
vibration-mediated resonant tunneling in the present model, obtaining the %%@
$I$-$V$ characteristics, zero-frequency current noise, and the effects of %%@
environmental dissipation on the role of the unequilibrated phonons.           
Our numerical analysis shows that in the voltage region where the ground %%@
orbital is dominant in tunneling, the combined effect of unequilibrated %%@
phonons and asymmetric tunnel-couplings is responsible for weak peaklike %%@
structures in the $I$-$V$ curve at the onset of each phonon step with a weak %%@
NDC and correspondingly weak super-Poissonian noise. Furthermore we found that %%@
this peaklike structure could be gradually diminished by environmental %%@
dissipation of the uneqilibrated IVM and become completely devoid of NDC at %%@
the equilibrated phonon condition. Accordingly, the usual current noise for an %%@
asymmetric single-electron tunneling device is predicted at the equilibrated %%@
phonon condition, $F=1$; however, for a {\em finite} dissipation rate %%@
suppressed noise, $F<1$, may be observed under certain conditions.   

More interestingly, we have also discussed the transport properties in detail %%@
in the second-orbital dominated region. We found that the interplay of strong %%@
Coulomb interaction between the two MOs and strong asymmetry of %%@
tunnel-coupling leads to very strong NDC and a correspondingly strongly %%@
enhanced shot noise, irrespective of the EPC. However, the bias voltage value %%@
for the onset of NDC and super-Poissonian shot noise is intimately dependent %%@
on the EPC strength of the second MO, i.e., this voltage value can be smaller %%@
or larger than the traditional resonant tunneling value for the second MO. Our %%@
discussion concluded that this feature stems from the EPC-induced selective %%@
cascades of single-electron transitions with FC-factor-modified rates in a %%@
unidirectional way due to the asymmetric tunnel-coupling. Environmental %%@
dissipation forces this value to tend to the traditional resonance point.

\begin{acknowledgments} 

This work was supported by Projects of the National Science Foundation of %%@
China, the Shanghai Municipal Commission of Science and Technology, the %%@
Shanghai Pujiang Program, and Program for New Century Excellent Talents in %%@
University (NCET). NJMH was supported by the DURINT Program administered by %%@
the US Army Research Office, DAAD Grant
No.19-01-1-0592.

\end{acknowledgments}


\begin{thebibliography}{99}
\bibitem{dyadic}{The dyadic operators are pseudo-operators defined as: ${\hat %%@
a}^\dagger=|A\rangle$ and $\hat a=\langle A|$ ($|A\rangle$ is a quantum %%@
state). The combination of dyadic operators gives a real operator ${\hat %%@
a}^\dagger b=|A\rangle\langle B|$ describing the transition from the state B %%@
to the state A, or the inner product $\hat b {\hat a}^\dagger=\langle %%@
B|A\rangle$ between the states $|A\rangle$ and $|B\rangle$.}
\end{thebibliography}
\end{document}